\documentclass{seg}
\usepackage{pgfplots}
\usepackage[letterpaper,top=2cm,bottom=2cm,left=3cm,right=3cm,marginparwidth=1.75cm]{geometry}
\usepackage{natbib} 

\title{Broadband Multi-Aperture Passive Scholte-Wave Imaging Using Seabed Distributed Acoustic Sensing\footnotemark[1]}

\author{%
  Anna Titova$^{1}$\and Andrey Bakulin$^{1}$}
  
\date{\today}

\begin{document}

\thispagestyle{empty} 
\vspace*{\fill} 
\noindent {\large This manuscript ``Broadband Multi-Aperture Passive Scholte-Wave Imaging Using Seabed Distributed Acoustic Sensing" is a non-peer reviewed preprint, which has been submitted to \textit{The Seismic Record} (Seismological Society of America) for peer review  on July 14, 2026. We encourage your feedback.\\

\noindent Anna Titova, \\
Bureau of Economic Geology, The University of Texas at Austin, Austin, 78758,  U.S.A.\\

\noindent Andrey Bakulin, \\
Bureau of Economic Geology, The University of Texas at Austin, Austin, 78758, U.S.A.\\

\noindent This preprint is under arXiv.org perpetual, non-exclusive license as the authors submitting a manuscript to \textit{The Seismic Record} are required to retain copyright and follow the Seismological Society of America policy, which implies that the preprint of the submitted manuscript should not be published under the Creative Commons license.}

\vspace*{\fill} 
\newpage 
\begin{titlepage} 
\maketitle 
\footnotetext[1]{Preprint version -- the original is submitted for peer review to \textit{The Seismic Record} (Seismological Society of America) on July 14, 2026. This preprint is under arXiv.org perpetual, non-exclusive license as the authors submitting a manuscript to \textit{The Seismic Record} are required to retain copyright and follow the Seismological Society of America policy, which implies that the preprint of the submitted manuscript should not be published under the Creative Commons license.}

\begin{center}
    \footnotesize
    $^{1}$Bureau of Economic Geology, The University of Texas at Austin, Austin, 75758  U.S.A.\\ {https://orcid.org/0000-0002-7759-2230}{(AT)};\\{https://orcid.org/0000-0002-6638-7821}{(AB)}
\end{center}

\begin{abstract}
Shallow-water ocean forcing provides strong broadband passive Scholte-wave illumination along a seabed distributed acoustic sensing cable on the Texas Gulf Coast. We exploit this illumination through direct processing of the uncorrelated wavefield, long-duration frequency-wavenumber stacking, and a multi-aperture strategy that preserves high-frequency spatial localization while resolving low-frequency modes. Constant-wavenumber slices enable spatially consistent multimode tracking across successive array centers. The resulting dispersion spans approximately $0.3$--$4.5$~Hz and supports 400 one-dimensional inversions along a $51$-km cable segment. These profiles form a pseudo-2D shear-wave velocity model extending from the near seafloor to approximately $2$~km depth. Broad shallow low-velocity intervals are consistent with softer incised-valley fill within stiffer Pleistocene deposits. The results demonstrate that existing seabed fiber infrastructure, when coupled with strong shallow-water passive illumination, can deliver broadband regional shear-wave imaging, while higher-frequency active sources remain necessary to resolve the uppermost several meters.
\end{abstract}
\end{titlepage}
\newpage
\section*{Introduction}
Shear-wave velocity is a key indicator of sediment stiffness, consolidation state, and stratigraphic variability across heterogeneous continental-shelf environments such as incised valleys, transgressive deposits, and channel systems \citep{simms2006overfilled,menier2006basement}. Surveys by the Texas General Land Office and the Bureau of Ocean Energy Management have mapped shallow sedimentary architecture and offshore sand resources using sub-bottom profiles, bathymetry, sidescan sonar, and localized vibracores \citep{aptim2025texas}. These observations constrain depositional geometry, seismic facies, sediment thickness, and near-seafloor composition, but they do not provide a spatially continuous regional shear-wave velocity model. Such a model would complement reflection and core observations by quantifying mechanical properties over a broader depth range and improving predictions of seismic-wave interaction with the seabed \citep{kennett1996does,liu2023modeling}.

Scholte waves are strongly controlled by marine-sediment shear-wave velocity and provide a practical means of estimating subsurface stiffness \citep{biot1952interaction,herrmann2013computer,long2020multichannel,wang2020seismic}. Active-source surveys can resolve the upper tens of meters but are limited by source deployment and acquisition effort. Passive Scholte waves provide longer wavelengths and regional coverage, while seabed distributed acoustic sensing (DAS) offers continuous, densely sampled measurements along long fiber-optic cables.

Extracting multimode dispersion from passive DAS records remains difficult because ambient forcing is uncontrolled, signal strength varies with frequency and location, and no single spatial aperture can both resolve low-frequency wavenumbers and preserve short-wavelength information in laterally heterogeneous sediments. Passive surface-wave methods may be interferometric, based on cross-correlations or spatial coherence, or non-interferometric, based on direct analysis of ambient records \citep{cheng2023comparisons}. Direct extraction from uncorrelated seabed DAS data was demonstrated by \cite{spica2020marine}, whereas correlation-based workflows were used by \cite{williams2021scholte} and \cite{cheng2021utilizing}.

Here, we process the original passive wavefield directly and stack frequency-wavenumber (FK) spectra over time to emphasize persistent dispersion for a static shear-wave velocity model. The shallow-water setting provides sustained ocean-generated energy across a broad band, consistent with local microseism generation by coastal interaction and reflection of gravity waves \citep{guerin2022quantifying}.

We develop a broadband workflow based on time-ensemble FK stacking, multi-aperture analysis, and spatially consistent dispersion tracking using wave-number slices. Direct processing preserves coherent Scholte-wave energy from approximately $0.3$ to $4.5$~Hz, whereas our ongoing correlation-based processing of the same dataset is largely restricted to frequencies below about $2$~Hz. Short apertures preserve high-fre-quency localization, long apertures improve low-frequency wavenumber resolution, and wavenumber slices maintain consistent multimode tracking across laterally variable areas where panel-by-panel picking may become ambiguous or fail. 

The workflow is applied to a 65-km seabed fiber-optic cable on the inner and middle Texas continental shelf. Four hundred shear-wave velocity profiles are estimated along an approximately 51-km segment and assembled into a pseudo-2D model extending from the near seafloor to kilometer-scale depths. The model reveals broad lateral and vertical variability, including shallow velocity contrasts consistent with transitions between stiffer Pleistocene deposits and softer incised-valley fill \citep{simms2006overfilled}. The study therefore contributes both a broadband passive imaging method and a regional shear-wave velocity product that complements existing geotechnical and reflection-seismic characterization of the Texas shelf.

\section*{Broadband Multi-Aperture Scholte-Wave Imaging}

The passive Scholte-wave field varies with time, frequency, sea state, and water depth. Consequently, the duration of temporal stacking required to obtain stable dispersion spectra is data dependent. Because the objective is a static shear-wave velocity model rather than time-lapse analysis, FK spectra are stacked over sufficiently long intervals to enhance persistent Scholte-wave energy.

\begin{figure}[h!]
\centering
\includegraphics[width=0.45\textwidth]{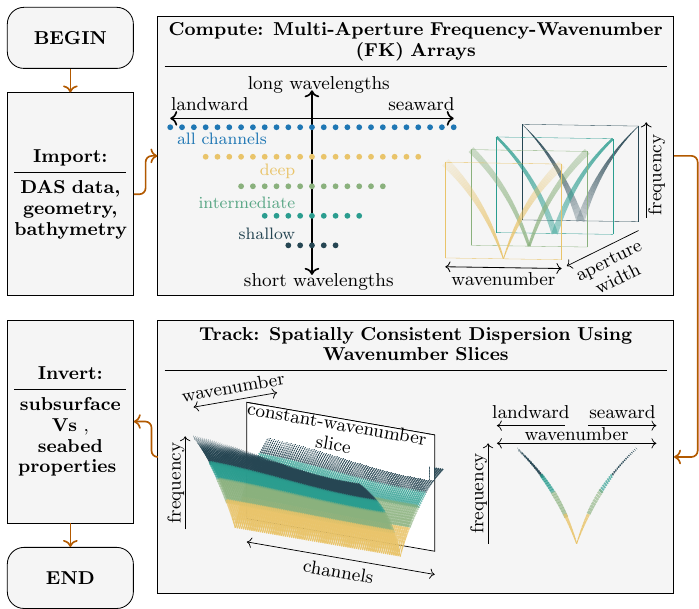}
\caption{Workflow for broadband passive Scholte-wave imaging using seabed DAS. Time-stacked FK spectra are computed over multiple spatial apertures to balance high-frequency spatial localization against low-frequency wavenumber resolution. Constant-wavenumber slices are then used for spatially consistent multimode dispersion tracking across laterally variable areas before inversion for shear-wave velocity. Negative and positive wavenumbers represent landward- and seaward-propagating energy, respectively.
}
\label{fig:workflow}
\end{figure}

For the same dataset, parallel correlation-based processing currently provides reliable dispersion primarily below approximately $2$ Hz, whereas direct processing of the uncorrelated passive wavefield retains coherent energy to approximately $4.5$ Hz. Preserving this broader band improves shallow sensitivity through the higher frequencies while maintaining deeper sensitivity through the lower frequencies. The workflow shown in Figure~\ref{fig:workflow} combines time-ensemble FK stacking, multi-aperture analysis, spatially consistent multimode dispersion tracking using wavenumber slices, and multimodal surface-wave inversion.

\subsection*{Multi-Aperture FK Arrays}

Non-interferometric processing of the passive DAS data is performed by applying a two-dimensional (2D) Fourier transform to the recorded DAS traces. Each ten-minute record is transformed independently, and the resulting FK spectral magnitudes are stacked over the acquisition period of 8 hours. FK analysis does not require knowledge of source location or timing because the transform maps spatially and temporally coherent wavefield components into frequency-wavenumber space. Stacking suppresses transient fluctuations and incoherent energy while enhancing persistent Scholte-wave dispersion. Hereafter, the time-stacked result is referred to simply as the FK spectrum. 

No trace-by-trace or spectrum-by-spectrum normalization is applied before or after stacking. This choice avoids artificially equalizing spatial amplitude variations and preserves relative amplitudes for quality control and potential future amplitude-derived attributes. The dispersion measurements used here depend on the locations of coherent FK maxima rather than on their absolute amplitudes.

A multi-aperture FK array assembles individual FK spectra calculated for different spatial aperture widths and referenced to the same central DAS channel. The aperture sequence is constructed by starting with the shortest aperture and progressively increasing it to the longest. Shorter apertures preserve high-frequency, short-wavelength dispersion where shallow structure varies laterally, whereas longer apertures improve wavenumber resolution and mode separation at low frequencies. This progression is conceptually similar to the transition from high to low frequencies used by \cite{masoni2016layer} in layer-stripping full-waveform inversion of surface waves. The field-data criteria used to select the aperture sequence are described below.

Because the dataset contains thousands of DAS channels and long continuous records, calculating several FK spectra for many overlapping arrays is computationally demanding. To reduce repeated operations, the 2D Fourier transform is represented as the Kronecker product of temporal and spatial one-dimensional transforms,
\begin{equation}
    \mathbf{F}_{tx} = \mathbf{F}_{x}^{H} \otimes \mathbf{F}_{t}^{H},
\end{equation}
where $\mathbf{F}_{t}$ and $\mathbf{F}_{x}$ denote the temporal and spatial Fourier operators, respectively \citep{titova2019mutual}. The temporal Fourier transform is therefore calculated once, stored in fast-access memory, and reused when evaluating the different spatial apertures.

\subsection*{Spatially Consistent Dispersion Tracking Using Wavenumber Slices}

In conventional FK-based surface-wave analysis, each dispersion image is calculated from a multichannel spatial aperture. Moving the aperture along the receiver line produces a sequence of local FK spectra referenced to successive central DAS channels. Dispersion is commonly interpreted independently in each local FK panel and subsequently assembled into a spatially varying dataset. This panel-by-panel procedure may lead to inconsistent mode identification when adjacent branches approach one another or different modes dominate at  neighboring locations.

Here, the sequence of local FK spectra is treated as a three-dimensional representation of spectral amplitude as a function of frequency, wavenumber, and FK-array center. Rather than picking dispersion independently in each frequency-wavenumber panel, this representation is sliced at constant wavenumber. Spectral values from successive FK-array centers are thereby arranged in a frequency-channel plane, referred to here as a wavenumber slice. The reorganization is illustrated schematically in Figure~\ref{fig:workflow}, and field-data examples are presented in the next section. Dispersion can then be tracked along the cable at a fixed wavenumber, using spatial continuity to maintain consistent mode identification.

Tracking begins at the largest absolute wavenumber and is performed separately for the landward- and seaward-propagating branches. Within each slice, local spectral maxima are identified by comparison with neighboring frequency samples. Upper and lower frequency bounds are assigned manually for each mode to prevent switching between adjacent branches. The same bounds can generally be applied across all FK-array centers within one slice, while their positions are adjusted monotonically between successive slices to follow the evolution of each mode. Corresponding negative and positive wavenumbers are interpreted using the same modal frequency bounds.

Here, $k$ denotes spatial frequency in cycles per meter rather than angular wavenumber, such that $\lambda=1/|k|$ and the phase-velocity magnitude is $v=f/|k|$.

\subsection*{Multimodal Scholte-Wave Inversion}

The dispersion picks are filtered to remove isolated outliers and inverted using the gradient-based multimodal surface-wave inversion implemented in \textit{Computer Programs in Seismology} \citep{herrmann2013computer}. This approach has been widely applied to Scholte-wave dispersion in horizontally layered marine models \citep{wang2020seismic}. In this study, the inversion uses dispersion measurements spanning approximately $0.3$--$4.5$~Hz.

A common initial model is used at all locations. Its shear-wave velocity profile is obtained by applying a pseudo-depth transformation to a
representative regional set of dispersion picks and fitting a smooth power-law trend. Sediment-layer thicknesses increase approximately logarithmically with depth, except within the upper $2$~m, where uniform $0.1$~m layers are used. Compressional-wave velocity and density are spatially uniform and held fixed during inversion at $1500$~m/s and $1030$~kg/m$^3$, respectively.

The water column is included explicitly as the uppermost acoustic layer, with its thickness determined from the bathymetry profile along the cable. Compressional-wave velocity and density are held fixed during inversion. Differential smoothing is applied to the sediment layers, and damping is progressively reduced through the inversion iterations. Dispersion curves are inverted independently at successive locations, and the resulting one-dimensional shear-wave velocity profiles are assembled into a pseudo-2D section.

\section*{Field Data Application}

\begin{figure}[h!]
\centering
\includegraphics[width=0.75\textwidth]{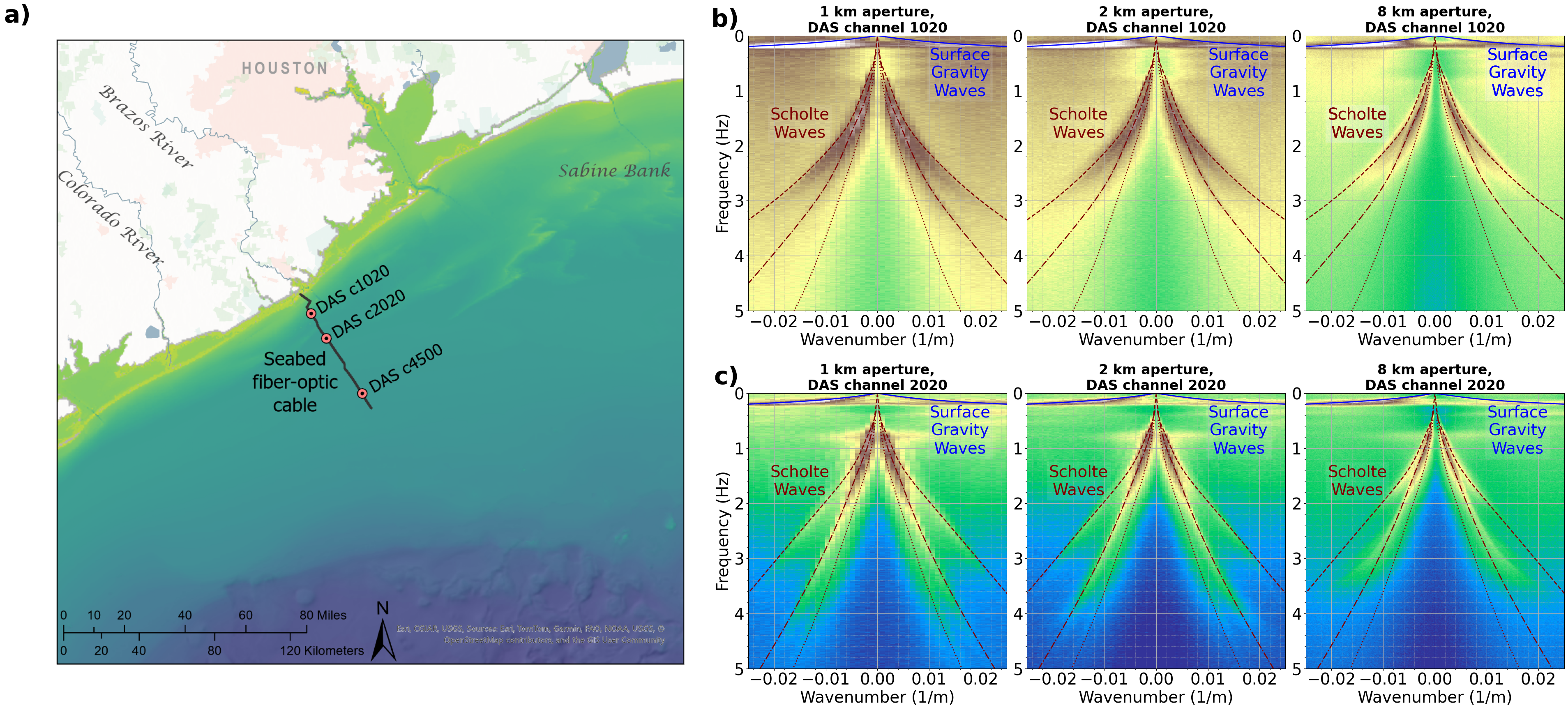}
\caption{Study area, seabed fiber-optic cable geometry, and representative time-stacked FK spectra. (a) Processed cable segment on the inner and middle Texas continental shelf. Markers indicate the locations of DAS channels 1020 and 2020. (b) FK spectra at DAS channel 1020, located approximately $13$ km along the cable from shore at a water depth of approximately $17$ m. (c) FK spectra at DAS channel 2020, located approximately $26$ km along the cable from shore at a water depth of approximately $21$ m. Spectra are calculated using nominal aperture lengths of $1.02$, $2.04$, and $8.2$ km. Modeled surface-gravity-wave and Scholte-wave dispersion curves are overlaid to guide interpretation. Increasing the aperture improves wavenumber resolution and mode separation at low frequencies, particularly below approximately $1.5$ Hz. At higher frequencies, longer apertures may reduce dispersion coherence by averaging wavefields over laterally varying shallow structure, as most clearly observed at DAS channel 2020. These examples motivate the use of multiple apertures to preserve the approximately $0.3$--$4.5$ Hz usable bandwidth. }
\label{fig:map_and_data}
\end{figure}

The study area covers the inner and middle Texas continental shelf west of Sabine Bank and southwest of Houston, Texas (Figure~\ref{fig:map_and_data}a). The telecommunication fiber-optic cable begins onshore near the modern Brazos River delta and extends seaward into the deeper Gulf basin. The dataset analyzed here spans one day, October 1, 2026. DAS acquisition used a gauge length of $23.93$~m and a channel spacing of $12.76$~m, producing 5,091 channels along approximately $65$~km of cable. The records were downsampled to $10$~Hz and organized into continuous ten-minute segments for FK processing. Along the shallow-water portion used in this study, the cable is buried approximately $1$~m beneath the seabed, and the DAS interrogator continuously records changes in axial strain.

Figure~\ref{fig:map_and_data}a shows the cable geometry and representative time-stacked FK spectra calculated at two locations using different spatial apertures. These examples demonstrate the central aperture trade-off underlying the broadband workflow: shorter apertures preserve locally varying high-frequency dispersion, whereas longer apertures improve the separation of low-frequency energy along the wavenumber axis.

\subsection*{Data-Driven Aperture Selection}

The aperture sequence is selected directly from the time-stacked field-data spectra rather than from a fixed aperture-to-wavelength ratio. The shortest member of the sequence is chosen as the maximum aperture length that preserves coherent high-frequency dispersion without significant lateral smearing. For the examples in Figure~\ref{fig:map_and_data}b and c, an aperture of approximately $1.02$~km preserves identifiable Scholte-wave energy to about $4.5$~Hz. Increasing the aperture averages the wavefield over a larger interval of laterally varying shallow structure and progressively broadens the high-frequency dispersion, particularly at the location shown in Figure~\ref{fig:map_and_data}c.

The $1.02$~km spectra are then examined at low frequencies to determine whether the corresponding wavenumber resolution is sufficient to separate the dispersion branches. Below approximately $1.5$~Hz, several modeled branches occupy the same or adjacent wavenumber bins and are therefore poorly resolved. Increasing the aperture reduces the wavenumber-bin spacing and improves separation of the low-frequency modes. The aperture sequence is extended until the lowest usable frequencies are adequately resolved. For the present dataset, the selected aperture lengths span approximately $1.02$ to $8.20$~km.

The lowest consistently usable frequency is approximately $0.3$~Hz, corresponding to the lower end of the locally generated microseism energy
observed in the shallow-water records. Energy is also present between approximately $0.2$ and $0.3$~Hz, but stable dispersion retrieval in this interval is complicated by spectral leakage from strong surface-gravity-wave energy. The selected aperture sequence therefore supports dispersion analysis over an effective frequency range of approximately $0.3$--$4.5$~Hz.

The selected aperture lengths are assigned to complementary spatial-frequency ranges when assembling the dispersion dataset.  The approximately $1.02$~km aperture is used for $0.008\leq|k|\leq0.02$~m$^{-1}$, the approximately $2.04$~km aperture for $0.0039\leq|k|\leq0.02$~m$^{-1}$, the approximately $4.08$~km aperture for $0.0024\leq|k|\leq0.0083$~m$^{-1}$, and the approximately $8.20$~km aperture for $0.00025\leq|k|\leq0.005$~m$^{-1}$. Adjacent ranges overlap near the transition spatial frequencies, where the aperture providing clearer modal separation is selected. The same aperture-to-spatial-frequency assignment is applied at all FK-array centers.

\subsection*{Wavenumber-Slice Dispersion Analysis}

The construction of a wavenumber slice is illustrated schematically in Figure~\ref{fig:workflow}. Local FK spectra centered at successive DAS channels form a three-dimensional representation of spectral amplitude as a function of frequency, wavenumber, and FK-array center. Extracting spectral values at a fixed wavenumber produces a frequency-channel section in which dispersion can be tracked continuously along the cable. 

\begin{figure}[h!]
\centering
\includegraphics[width=0.60\textwidth]{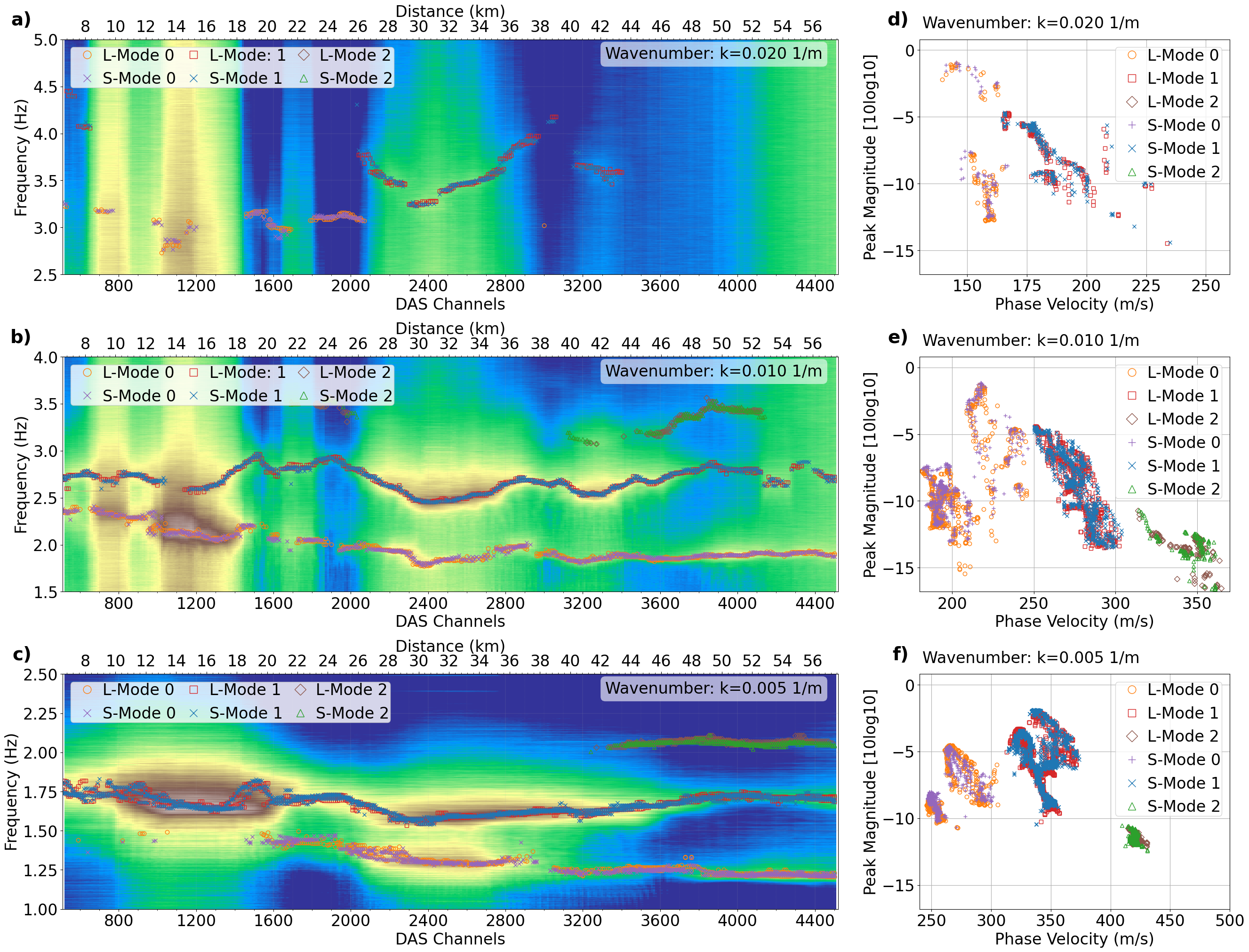}
\caption{Field-data wavenumber slices formed by extracting constant-wavenumber values from the sequence of local FK spectra. Interpreted multimodal dispersion is shown at (a) $k=0.020$~m$^{-1}$, (b) $k=0.010$~m$^{-1}$, and (c) $k=0.005$~m$^{-1}$. Panels (d)--(f) show the amplitudes of the picked spectral maxima as a function of phase-velocity magnitude, $v=f/|k|$. Symbols distinguish the interpreted modes and the independently picked landward- and seaward-propagating branches.}
\label{fig:wavenumberslices}
\end{figure}

Figure~\ref{fig:wavenumberslices} shows field-data examples at three selected wavenumbers spanning more than $50$~km of the cable. The fundamental mode dominates in some intervals, whereas the first overtone becomes stronger elsewhere. Despite these changes in relative modal strength, the spatial continuity of the dispersion trends allows the modes to be followed consistently without interpreting each local FK panel independently.

Figure~\ref{fig:wavenumberslices}d--f provides a complementary quality-control view of the dispersion interpretation. Each point represents the amplitude of a picked spectral maximum from the corresponding wavenumber slice, plotted against phase-velocity magnitude calculated as $v=f/|k|$. Because the channel coordinate is not retained in these panels, they do not show spatial continuity directly. Instead, they summarize the separation of the interpreted modal populations. Separation between phase-velocity clusters supports consistent mode assignment, whereas overlapping or diffuse clusters indicate greater ambiguity. The similar distributions obtained for the independently interpreted landward and seaward branches provide an additional check on the stability of the dispersion picks.

Wavenumber slices should not be interpreted as literal fixed-depth cross-sections. However, because the depth sensitivity of Scholte waves scales with horizontal wavelength \citep{stein2009introduction,foti2018guidelines}, the slices provide scale-selective views of lateral variability. Shorter wavelengths emphasize shallower structure, whereas longer wavelengths sample progressively greater depths.

At $k=0.020$~m$^{-1}$, corresponding to a wavelength of approximately $50$~m, Figure~\ref{fig:wavenumberslices}a shows a rapid change in the
short-wavelength dispersion near DAS channel 2000. At $k=0.010$~m$^{-1}$, corresponding to a wavelength of approximately $100$~m, reduced phase velocities and convergence of the fundamental mode and first overtone occur near DAS channel 800. At the longer wavelength of approximately $200$~m in Figure~\ref{fig:wavenumberslices}c, a spatially localized variation in the first overtone remains visible near the same portion of the cable. These features remain after dispersion-pick quality control but should be interpreted cautiously because strong lateral variability may violate the local one-dimensional assumption used in the subsequent inversion.

\subsection*{Shear-Wave Velocity Structure}

One-dimensional shear-wave velocity profiles were inverted at every tenth DAS channel, yielding 400 profiles at an output spacing of $127.6$~m along an approximately $51$-km-long cable segment. The same FK-array center spacing is used for all aperture lengths to keep the dispersion measurements spatially co-registered. The $1.02$- and $8.20$-km apertures overlap between neighboring centers by approximately $87.5\%$ and $98.4\%$, respectively. Consequently, the $127.6$~m spacing represents the sampling of the displayed section rather than its effective lateral resolution, which becomes progressively coarser toward lower frequencies and longer apertures. Only centers with a complete aperture sequence are retained, corresponding to DAS channels 500 through 4500.

\begin{figure}[h!]
\centering
\includegraphics[width=0.70\textwidth]{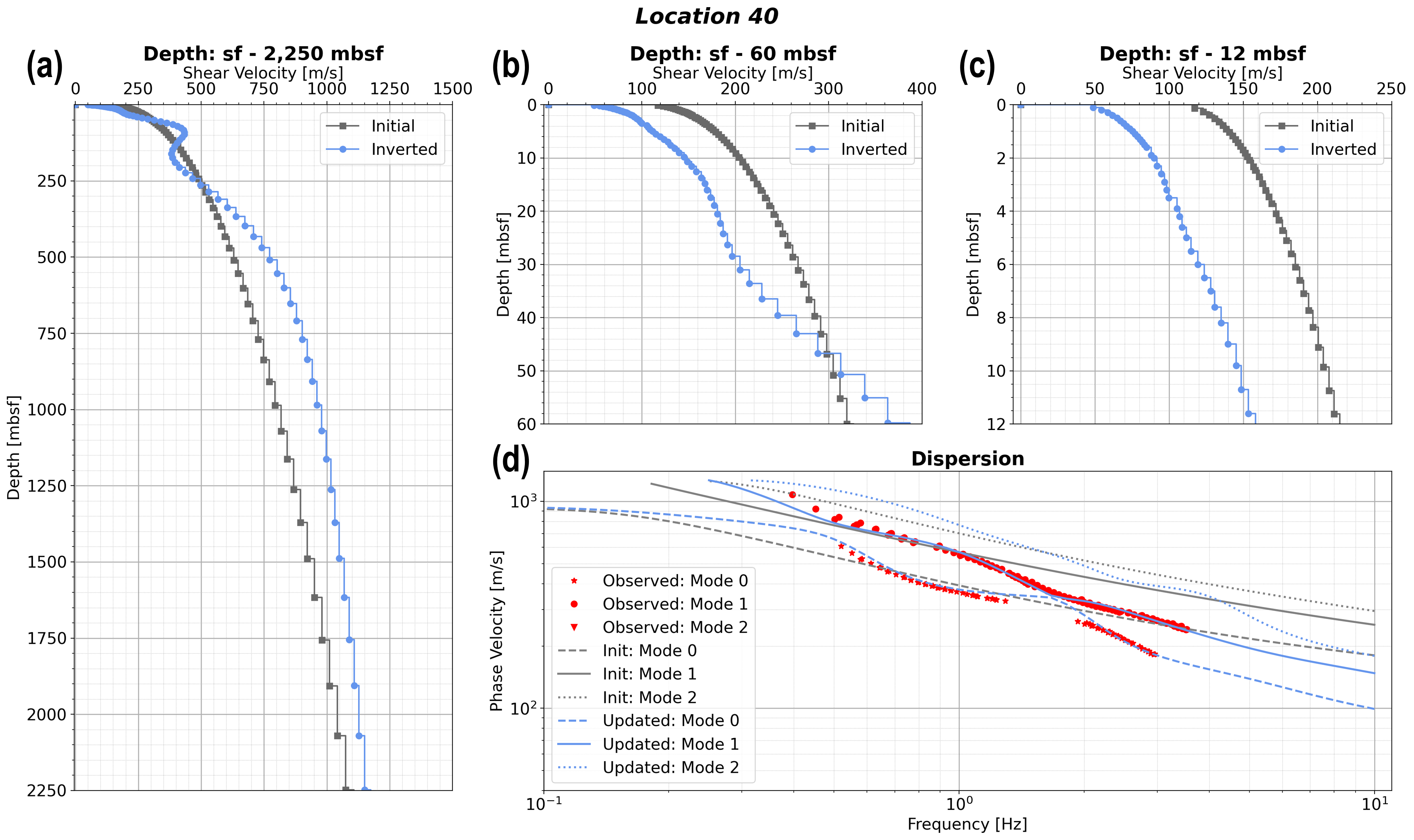}
\caption{Representative one-dimensional Scholte-wave inversion at location 40. Panels (a)--(c) show the same initial and inverted shear-wave velocity models over the full model depth, the upper $60$~m, and the upper $12$~m below the seafloor, respectively. The smooth initial model follows a power-law velocity trend. Layer thicknesses increase approximately logarithmically with depth, with uniform $0.1$~m layers used within the upper $2$~m. Panel (d) compares the observed multimode dispersion picks with theoretical curves predicted by the initial and inverted models. At this location, the inversion produces a pronounced low-velocity zone at approximately $100$--$250$~m depth that is 
absent from the initial model, while improving the multimode dispersion fit.}
\label{fig:example_inversion}
\end{figure}

Figure~\ref{fig:example_inversion} shows a representative inversion from this set. Starting from the common smooth power-law model, the inversion develops a pronounced low-velocity zone at approximately $100$--$250$~m depth while substantially improving the multimode dispersion fit. Comparable velocity reversals occur only locally along the cable. 

\begin{figure}[h!]
\centering
\includegraphics[width=0.7\textwidth]{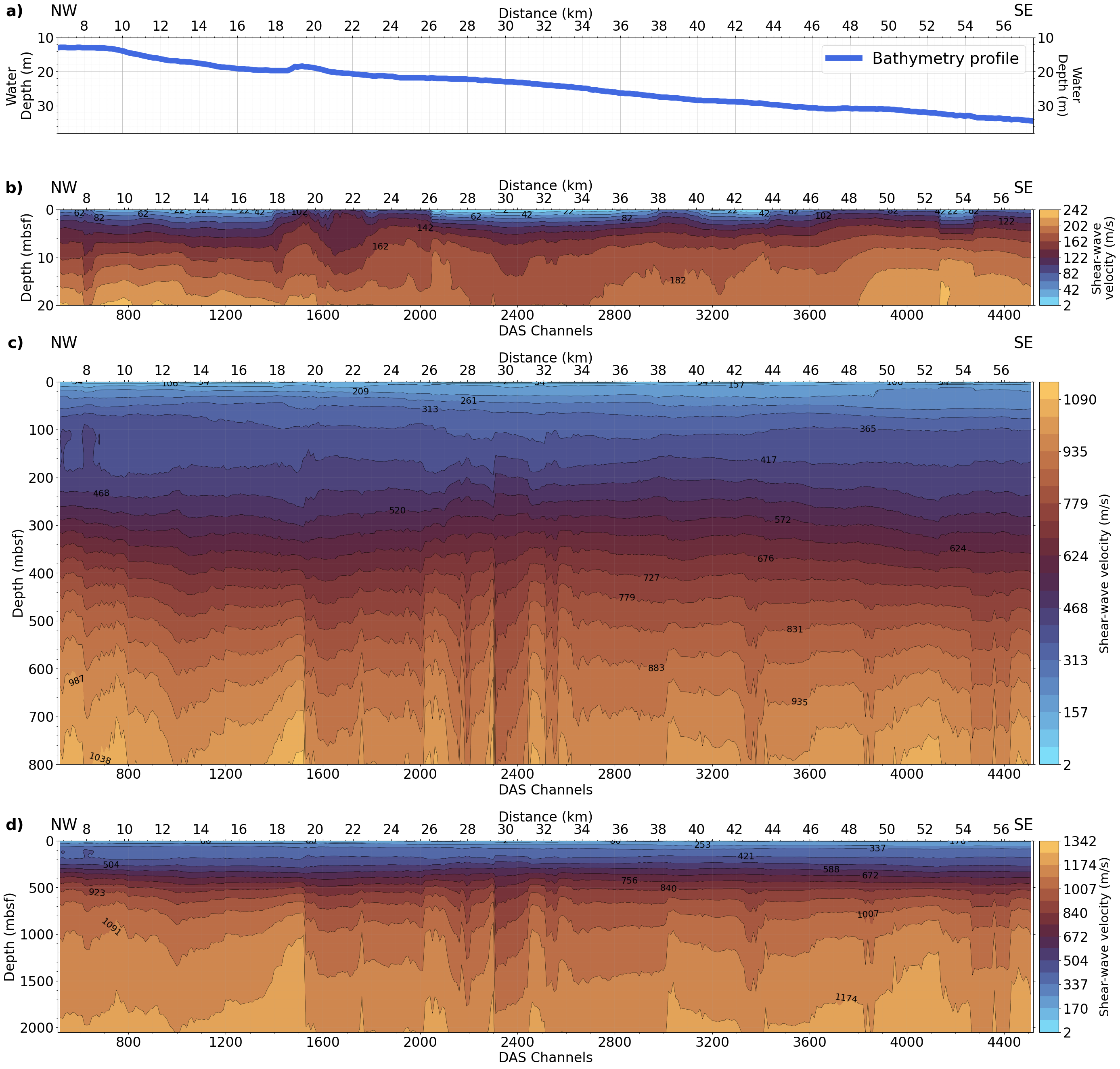}
\caption{Bathymetry and pseudo-2D shear-wave velocity structure along the processed cable segment. (a) Water depth increases from approximately $10$ to $36$~m from northwest (NW) to southeast (SE). The velocity model is displayed over depth intervals of (b) $0$--$20$~m below seafloor, (c) $0$--$800$~m below seafloor, and (d) $0$--$2000$~m below seafloor. The section is assembled from 400 independently inverted one-dimensional profiles sampled every $127.6$~m. Because the dispersion measurements use overlapping apertures of approximately
$1.02$--$8.20$~km, the profile spacing should not be interpreted as the effective lateral resolution. The passive bandwidth provides sensitivity to the upper tens of meters, but panel (b) represents broad near-seafloor velocity trends; fine-scale structure within the uppermost several meters is not independently resolved. The deepest portion of the model is also less well constrained and is interpreted primarily in terms of broad spatial trends.}
\label{fig:seabedprofiles}
\end{figure}

Figure~\ref{fig:seabedprofiles}a shows the bathymetry used to define the water layer in the inversion. Water depth increases gradually from approximately $10$ to $36$~m from northwest to southeast. Figures~\ref{fig:seabedprofiles}b--d show the resulting pseudo-2D shear-wave velocity section over three depth ranges. Each profile is inverted independently, and the 400 profiles are subsequently assembled into the displayed section without joint lateral regularization.

The upper $20$~m below the seafloor is enlarged in Figure~\ref{fig:seabedprofiles}b to show broad lateral variations in the near-seafloor model. The data are sensitive to this depth interval, but the highest usable frequencies of approximately $4$--$4.5$~Hz and the shortest measured wavelength of about $50$~m limit vertical resolution in the immediate near surface. In particular, fine-scale structure and sharp velocity contrasts within the uppermost several meters are not independently resolved. Panel \ref{fig:seabedprofiles}b should therefore be interpreted as a smoothed representation of broad near-seafloor velocity trends. 

The shallow section contains broad, kilometer-scale low-velocity intervals that locally deepen relative to the surrounding higher-velocity structure. Their dimensions and overall geometry are consistent with transitions between softer incised-valley fill and stiffer Pleistocene deposits previously mapped on the Texas continental shelf \citep{simms2006overfilled}.

Between approximately $50$ and $800$~m below the seafloor, Figure~\ref{fig:seabedprofiles}c shows a general increase in shear-wave velocity with depth, superimposed on moderate lateral variability. Near the northwestern end of the profile, a localized low-velocity region at approximately $100$--$250$~m depth includes the velocity reversal illustrated by the representative inversion in Figure~\ref{fig:example_inversion}. The cross-sectional view shows that this feature is laterally restricted, together with several broader velocity variations farther seaward.

Below approximately $800$~m, the increasing layer thickness and limited low-frequency sampling reduce vertical resolution. The model in Figure~\ref{fig:seabedprofiles}d therefore constrains mainly the long-wavelength increase in shear-wave velocity and broad lateral variations rather than detailed vertical structure. Values between approximately $1000$ and $2000$~m should be regarded as coarse regional estimates rather than resolved stratigraphic features.

\section*{Discussion}

The principal advantage of the workflow is preservation of a broad usable Scholte-wave band. Direct processing and long-duration FK stacking retain coherent energy from approximately $0.3$ to $4.5$~Hz, whereas the current correlation-based products from the same dataset are largely restricted to frequencies below about $2$~Hz. The additional high-frequency energy improves sensitivity in the upper tens to hundreds of meters, while the low-frequency end retains sensitivity to deeper structure. The bandwidth remains insufficient to resolve the uppermost few meters at geotechnical scale.

No single spatial aperture recovers this band optimally. Long apertures improve low-frequency wavenumber resolution and mode separation but average over larger lateral distances and may smear short-wavelength dispersion. Short apertures preserve high-frequency energy and local heterogeneity but provide inadequate resolution at low frequencies. The multi-aperture strategy therefore treats aperture as a scale-dependent processing parameter selected according to signal quality, wavelength, and expected lateral variability.

Direct passive-wavefield analysis also depends on source geometry. Strongly oblique plane waves recorded by a linear array produce apparent velocities greater than the true phase velocity and may bias dispersion estimates \citep{cheng2023comparisons}. Coastal wave interaction can instead generate repeated, spatially distributed microseism sources, including localized seismo-acoustic events associated with breaking surf \citep{guerin2022quantifying,francoeur2025identification}. Such sources may illuminate the cable over a range of azimuths rather than from a single ocean-wave direction. The persistence of very low apparent Scholte-wave velocities at high frequencies suggests that strongly oblique arrivals do not dominate the present measurements. Explicit source-direction analysis remains future work.

The resulting model complements, rather than replaces, existing GLO and BOEM bathymetric, sub-bottom, and geotechnical characterization. It adds a continuous regional shear-wave velocity section from the shallow marine interval to kilometer-scale depths. Localized velocity reversals within this section may reflect lithologic variation, underconsolidation, or excess pore pressure, but their origin cannot be distinguished from the present data. Such models may also improve elastic imaging, converted-wave interpretation, full-waveform inversion, and time-lapse monitoring because near-seabed shear stiffness influences energy partitioning among PP reflections, converted waves, and interface waves \citep{titova2026prestack}.

The near-seafloor resolution gap can be addressed with active sources. Seabed shear-wave vibrators provide excellent broadband multimode data but are operationally costly \citep{vanneste2011use}. A very small airgun towed a few meters above the seabed can recover Scholte-wave dispersion to approximately $20$~Hz and offers a practical complement to regional passive imaging \citep{trafford2022distributed}.

\section*{Conclusions}

Passive Scholte waves recorded by a seabed distributed acoustic sensing cable were used to construct a regional shear-wave velocity model of the inner and middle Texas continental shelf. Direct processing of the uncorrelated wavefield, combined with time-ensemble FK stacking, preserved coherent dispersion over approximately $0.3$--$4.5$~Hz. Multi-aperture analysis balanced high-frequency spatial localization against low-frequency wavenumber resolution, while constant-wavenumber slices enabled spatially consistent multimode tracking across successive FK-array centers.

The workflow produced 400 one-dimensional inversions along an approximately $51$-km cable segment, assembled into a pseudo-2D shear-wave velocity section. The model images broad lateral and vertical variations from the near seafloor to kilometer-scale depths. Shallow low-velocity intervals are consistent in scale and geometry with softer incised-valley fill embedded within stiffer Pleistocene deposits. Localized velocity reversals at approximately $100$--$250$~m depth provide an additional indication of subsurface complexity, although their geological origin remains uncertain.

The preserved passive bandwidth improves shallow sensitivity relative to correlation-based processing of the same dataset while retaining sensitivity to deeper structure. Broad velocity trends are obtained near the seafloor, but fine-scale structure within the uppermost several meters requires additional higher-frequency constraints. Existing seabed fiber-optic infrastructure can therefore support broadband passive imaging of marine shear-wave velocity over regional distances.

\section*{Data and Resources}
The data used in this study are not publicly available.

\section*{Declaration of Competing Interests}
The authors declare no competing interests.

\section*{Acknowledgements}
None.

\bibliographystyle{plain} 
\bibliography{references}

\end{document}